\documentclass[aps,prl,twocolumn,superscriptaddress]{revtex4}
\usepackage{amssymb}
\usepackage{amsfonts}
\usepackage{xcolor}
\usepackage{amsmath}
\usepackage{graphicx}
\usepackage{bm}
\begin{document}

\title{Resonant Plasmon-Assisted Tunneling in a Double Quantum Dot

Coupled to a Quantum-Hall Plasmon Resonator
}
\author{Chaojing Lin}
\email{lin.c.ad@m.titech.ac.jp}
\affiliation{Department of Physics, Tokyo Institute of Technology, 2-12-1 Ookayama,
Meguro, Tokyo, 152-8551, Japan.}
\affiliation{JST, PRESTO, 4-1-8 Honcho, Kawaguchi, Saitama 332-0012, Japan.}
\author{Ko Futamata}
\affiliation{Department of Physics, Tokyo Institute of Technology, 2-12-1 Ookayama,
Meguro, Tokyo, 152-8551, Japan.}
\author{Takafumi Akiho}
\affiliation{NTT Basic Research Laboratories, NTT Corporation, 3-1 Morinosato-Wakamiya,
Atsugi, 243-0198, Japan.}
\author{Koji Muraki}
\affiliation{NTT Basic Research Laboratories, NTT Corporation, 3-1 Morinosato-Wakamiya,
Atsugi, 243-0198, Japan.}
\author{Toshimasa Fujisawa}
\affiliation{Department of Physics, Tokyo Institute of Technology, 2-12-1 Ookayama,
Meguro, Tokyo, 152-8551, Japan.}

\begin{abstract}
Edge magnetoplasmon is an emergent chiral bosonic mode promising for studying electronic quantum optics. While the plasmon transport has been investigated with various techniques for decades, its coupling to a mesoscopic device remained unexplored. Here, we demonstrate the coupling between a single plasmon mode in a quantum Hall plasmon resonator and a double quantum dot (DQD). Resonant plasmon-assisted tunneling is observed in the DQD through absorbing or emitting plasmons stored in the resonator. By using the DQD as a spectrometer, the plasmon energy and the coupling strength are evaluated, which can be controlled by changing the electrostatic environment of the quantum Hall edge. The observed plasmon-electron coupling encourages us for studying strong coupling regimes of plasmonic cavity quantum electrodynamics.

\end{abstract}

\date{\today }
\maketitle


Quantum Hall (QH) edge channels provide unique opportunities for studying electronic quantum optics under strong Coulomb interactions in two complementary regimes \cite{Grenier11,Bocquillon14,Fujisawa22}. In the single-electron transport regime with flying electrons well isolated from the cold Fermi sea, the Coulomb repulsion between flying electrons induces correlated single-electron transport, as demonstrated in two-particle interferometry and quantum tomography experiments \cite{Bocquillon13,Fletcher23,Ubbelohde23,Ota19,Wang23,Rodriguez20,Freise20,Suzuki23,Jullien14,Bisognin19}. This potentially provides a mechanism for correlated quantum state transport from one functional device to another. In the plasmon transport regime with collective motion of the cold Fermi sea, the Coulomb interaction plays an essential role in formation of emergent bosonic modes from fermions, such as edge magnetoplasmon modes and Tomonaga-Luttinger liquids \cite{Chang03,Wen90,Grodnensky91,Bocquillon213,Inoue14,Kamata14,Hashisaka17,Hashisaka21,Hashisaka18}. While the classical wave nature of edge plasmons is clearly resolved, for example, with plasmon resonators and interferometers \cite{Ashoori92,Hiyama15,Kumada214,Talyanskii92}, the quantum plasmon mode is revealed recently \cite{Bartolomei23}. Application of these electronic modes in either regime to a mesoscopic functional device should pave the way for exploring electronic quantum optics in QH systems known as a 2D topological insulator.

The edge plasmons have several advantages for this direction. Thanks to the chirality of the QH system, plasmons as well as flying electrons travel unidirectionally without showing backscattering \cite{Hashisaka18}. The plasmon wavelength (about 100 $\mu$m at frequency of 4 GHz) is convenient for confining plasmons in a small region \cite{Talyanskii92,Kamata10}. Of particular interest is the high impedance $Z$ of the plasmon mode \cite{Bosco19}, which provides large voltage amplitude ($\widetilde{V}\simeq \sqrt{Zhf}$) for a given plasmon energy ($hf$). This is attractive for making a strong electric dipole coupling to external atoms or qubits with the scheme of cavity quantum electrodynamics (cQED) \cite{Devoret07,Childress04}. Remarkably, $Z = h/\nu e^2$ is quantized as a result of the topological state. One can reach significantly large $Z$ with small Landau-level filling factor $\nu$ = 2, 1, 1/3 $\ldots$, as compared to the impedance of the vacuum ($\simeq$ 377 $\Omega$) and state-of-the-art high-impedance transmission lines made of a Josephson junction array (a few kilohms) \cite{Altimiras13,Stockklauser17,Scarlino22}. Therefore, the edge plasmons are attractive for studying the strong, ultrastrong, and deep-strong coupling regimes of plasmonic cQED systems \cite{Bosco219}. These fascinating characteristics have stimulated theoretical proposals, such as long-range entanglement \cite{Elman17}. However, experimental handling of chiral plasmons remains challenging.

Here, we exploit a coupled plasmon-electron system, in which an on-chip QH plasmon resonator is capacitively coupled to a double quantum dot (DQD). Discrete resonant frequencies of the resonator manifest the wave nature of plasmons, and the resonance allows us to access a single plasmon mode with a fixed energy. The coupling induced resonant plasmon assisted tunneling (PlAT) in the DQD is observed, where the tunneling is allowed by absorption and emission of single or multiple plasmons. We clarify the plasmon-electron coupling based on the unique frequency and magnetic-field dependence of the PlAT. Some prospects for cQED applications are described. These results encourage us to study strong coupling regimes.

\begin{figure}
\includegraphics[width=8.5 cm]{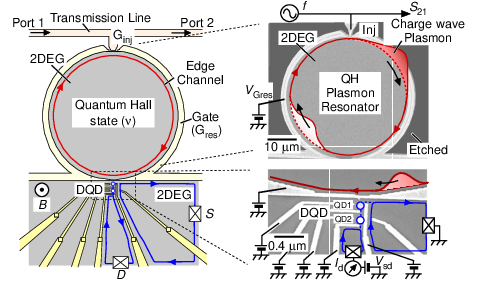}
\caption{Schematic of the device (left) and SEM pictures of the plasmon resonator (upper right) and the DQD (lower right). The plasmon resonator is formed as a circular QH channel propagating along the inner edge of the ring-shaped gate electrode G$_{\mathrm{res}}$. The upper half of the 2DEG outside the resonator (the black region in the upper SEM) has been removed by wet etching. The resonant mode is represented by wave packets with excess and deficit charges. A gate defined DQD (the two small circles in the lower SEM) is fabricated in the vicinity of the resonator and capacitively coupled to the resonator.}
\end{figure}

We propose and demonstrate a novel hybrid system consisting of a plasmon resonator in an isolated circular QH region and a charge two-level system (TLS) in a DQD, as shown in Fig. 1. The device was fabricated in a GaAs/AlGaAs heterostructure with a 2D electron gas (2DEG) having an electron density of $1.8\times10^{11}$ cm$^{-2}$ (see Supplemental Material (SM) \cite{SM}). Under a magnetic field $B$ perpendicular to the 2DEG, the plasmon resonator is defined by applying negative voltage $V_{\mathrm{Gres}}$ ($< -$0.28 V) to the ring-shaped gate G$_{\mathrm{res}}$ with perimeter $L$ = 126 $\mu$m (the resonant frequency $f_0$ $\sim$ 3 GHz at $\nu$ = 2). The plasmon mode (charge density wave) can be excited by applying microwave voltage to the top electrode G$_{\mathrm{inj}}$ through the coplanar transmission line and with a capacitive coupling scheme as shown in Fig. 2(a). We investigate RF transmission coefficient $S_{21}$ from port 1 to port 2 of the coplanar line to characterize the resonator. The plasmon resonance can be simulated by a distributed circuit model \cite{SM}, as shown in Fig. 2(b). A gate defined DQD was formed in the vicinity of the plasmon resonator. A TLS can be defined by choosing a single level in each dot. The static energy bias $\varepsilon$ and the tunneling coupling $t_{12}$ of the DQD can be varied by tuning the gate voltages. The tunneling current $I_{\mathrm{d}}$ from the source ($S$) through the DQD to the drain ($D$) was measured with applying a source-drain voltage $V_{\mathrm{sd}}$. The DQD is tuned in a weak tunnel coupling regime ($t_{12} \ll hf_0$), where clear PlAT is expected with the large permanent dipole moment. All measurements are performed at a temperature of $\sim$85 mK.

\begin{figure}[t]
\includegraphics[width=8.5 cm]{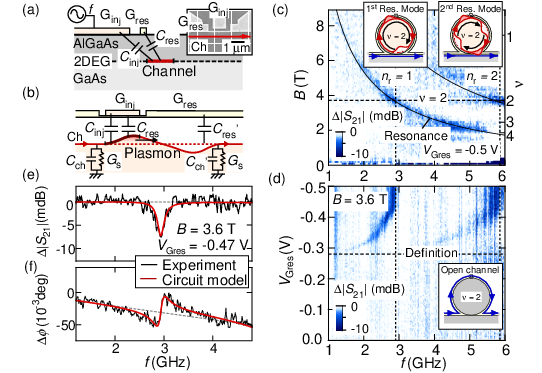}
\caption{(a) Schematic cross section of the edge channel and the injection gate G$_{\mathrm{inj}}$. A top-view SEM picture is shown in the inset. (b) Distributed circuit model for the edge channel (Ch) with coupling capacitances, $C_{\mathrm{ch}}$, $C_{\mathrm{res}}$, and $C_{\mathrm{inj}}$ in the injection region, $C_{\mathrm{ch}}'$ and $C_{\mathrm{res}}'$ in the other region, and dissipative conductance $G_{\mathrm{s}}$ in both regions. (c) Transmission spectra $\Delta|S_{21}|(f, B)$ obtained at $V_{\mathrm{Gres}} = -$0.5 V. The right axis shows the filling factor $\nu$. The resonant plasmon signals are seen in the blue regions for $n_{\mathrm{r}} $ = 1 and 2 with the charge density profiles illustrated in the insets. (d) $\Delta|S_{21}|(f, V_{\mathrm{Gres}})$ measured at $B$ = 3.6 T. (e) The amplitude $\Delta|S_{21}|$ and (f) the phase $\Delta\phi$ obtained at $B$ = 3.6 T and $V_{\mathrm{Gres}} = -$0.47 V. The red lines are calculated with distributed circuit model.}
\end{figure}

First, we evaluate the plasmon resonator by measuring $S_{21}$ with the resonator activated at negative $V_{\mathrm{Gres}}$, while other gates for the DQD were grounded. The resonance should appear in tiny change in $|S_{21}|$, which can be visible by subtracting the background signal $|S_{21}^0|$ obtained at $V_{\mathrm{Gres}}$ = 0 V without forming the plasmon resonator. Figure 2(c) shows the subtracted spectra $\Delta|S_{21}|$ = $|S_{21}| - |S_{21}^0|$ (in the decibel unit) as a function of frequency $f$ and magnetic field $B$. The plasmon resonances are identified as negative peaks with $\Delta|S_{21}|$ $<$ 0 dB (the blue regions). The resonant frequency decreases monotonically with increasing $B$, which is the signature of the edge-magnetoplasmon mode \cite{Talyanskii92}. Considering that the plasmon velocity $v = \sigma_{xy}/C_{\Sigma}$ is determined by the Hall conductance $\sigma_{xy} = \nu e^2/h$ and the channel capacitance $C_{\Sigma}$, the resonant frequency of the $n_r$-th mode follows $f = n_rv/L$ \cite{Hashisaka13,Ota18}. This reproduces the resonant frequencies, as shown by the solid lines for $n_r$ = 1 and 2 with $C_{\Sigma}$ = 0.21 nF/m. This $C_{\Sigma}$ is consistent with previous studies on gate-defined edge channels \cite{Kamata10,Lin21}. The observed multiple resonant modes clearly manifest the wave nature of the plasmons. The $n_r$ = 1 resonance is significantly enhanced at $B$ $\simeq$ 3.6 T, where a $\nu$ = 2 QH state is formed with negligible longitudinal resistance. The resonance signal weakens as $\nu$ deviates from 2, possibly due to the dissipation attributed to local disorders around the channel or in the bulk \cite{Ashoori92,Lin221}. While charge and spin collective modes are anticipated at $\nu$ = 2 \cite{Hashisaka17}, we focus on the charge (edge-magnetoplasmon) mode with large $Z$ in this study.

The resonant frequency can be tuned electrostatically, as shown in the $V_{\mathrm{Gres}}$ dependence of $\Delta|S_{21}|$ at $B$ = 3.6 T in Fig. 2(d). The resonant peaks for both $n_r$ = 1 and 2 modes appear at sufficiently negative $V_{\mathrm{Gres}}$ below the definition voltage ($\sim -0.28$ V) and show blueshifts with more negative $V_{\mathrm{Gres}}$. This frequency shifts can be understood by considering the $V_{\mathrm{Gres}}$ dependence of the channel capacitance $C_{\Sigma}$, which decreases as $V_{\mathrm{Gres}}$ decreases and the edge channels move away from G$_{\mathrm{res}}$ \cite{Kamata10}. In the following experiments for PlAT, we focus on the plasmon mode at $B$ = 3.6 T and $V_{\mathrm{Gres}} = -0.47$ V, where the resonant frequency is $f_0$ = 2.95 GHz. The spectra of $\Delta|S_{21}|$ and $\Delta\phi$ = arg($S_{21})$ $-$ arg($S_{21}^0$) at this condition are plotted in Figs. 2(e) and 2(f), respectively. They are well reproduced by a model calculation (the red lines, described in SM \cite{SM}) with quality factor $Q$ = 18.

\begin{figure}[t]
\includegraphics[width=8.5 cm]{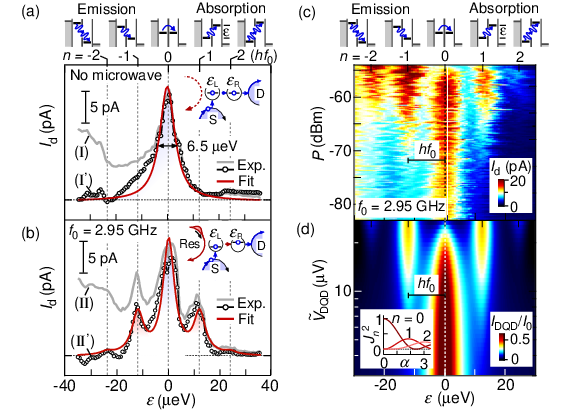}
\caption{(a-b) Current spectra $I_{\mathrm{d}}$ measured (a) at $P$ = 0 and (b) under plasmon irradiation at $P$ = -68 dBm. Traces (I) and (II) show the raw data. Traces (I') and (II') are obtained by removing the background current and fitted with a theoretical model (the red lines). (c) Current spectra measured at $f_0$ = 2.95 GHz for various RF powers. Satellite peaks are observed up to $|n|$ = 2 plasmons. (d) Calculated current spectra as a function of the ac voltage amplitude $\widetilde{V}_{\mathrm{DQD}}$ in the DQD. The inset shows the squared Bessel function $J_n^2(\alpha)$ of $\alpha = e\widetilde{V}_{\mathrm{DQD}}/hf_0$. The PlAT is illustrated in the upper insets of (a) and (c).}
\end{figure}

We now investigate the charge transport in the DQD with and without plasmon excitation. The DQD shows Coulomb charging energy of $\sim$0.8 meV, single-particle energy spacing of $\sim$140 $\mu$eV, and electrostatic coupling energy of $\sim$86 $\mu$eV, as shown in SM \cite{SM}. Trace (I) in Fig. 3(a) shows the tunneling current $I_{\mathrm{d}}$ through particular energy levels, $\varepsilon_L$ in the left dot and $\varepsilon_R$ in the right dot, measured at zero microwave power ($P$ = 0) and $V_{\mathrm{sd}}$ = 400 $\mu$V during the sweep of the gate voltages. The horizontal axis represents the energy detuning $\varepsilon$ = $\varepsilon_R$ - $\varepsilon_L$, where the detuning energy from the peak ($\varepsilon$ = 0) is obtained from the gate voltage with a lever-arm factor. For clarity, the background current associated with other nearby levels is subtracted from trace (I) (See SM \cite{SM}), and the subtracted trace (I') with open circles can be fitted by using a Lorentzian curve (the red line) with full width of $w_{\mathrm{DQD}}$ = 6.5 $\mu$eV. The deviation at $\varepsilon < $ 0 can be associated with the spontaneous phonon emission process \cite{Fujisawa98}.

\begin{figure}
\includegraphics[width=8. cm]{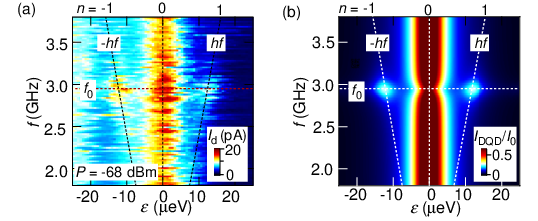}
\caption{(a) $f$ dependence of current spectra measured at $P = -$68 dBm. Satellite peaks appear only at the plasmon resonant frequency $f_0$ = 2.95 GHz. (b) Calculated current spectra.}
\end{figure}

The PlAT is observed under plasmon excitation, as shown by trace (II) of Fig. 3(b) obtained at $f_0$ = 2.95 GHz and microwave power $P = -$68 dBm estimated at the coplanar line (see SM) \cite{SM}. The background current from other levels is subtracted in trace (II'). The two distinct satellite peaks on both sides of the original peak are associated with absorption (the right peak at $\varepsilon$ $>$ 0) and emission (the left peak at $\varepsilon$ $<$ 0) of a single plasmon. This absorption tunneling is analogous to the photoelectric effect in which an electron is excited from a material by absorbing a photon. These peaks appear at $\varepsilon$ = $\pm$12 $\mu$eV, which coincides with the plasmon energy of $hf_0$ = 12.2 $\mu$eV. This means that a single plasmon can be removed from or added to the resonator by electron tunneling between the two dots. Significantly, these processes can be used for cQED applications.

\begin{figure*}[t]
\includegraphics[width=16 cm]{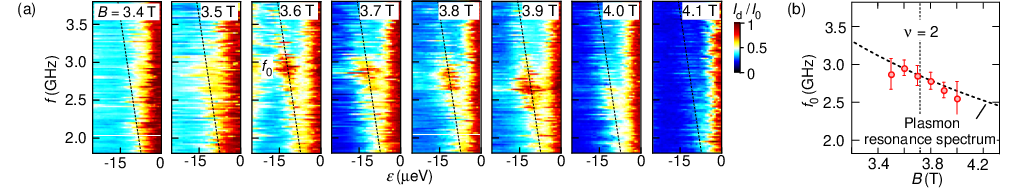}
\caption{(a) Current spectra measured at $B$ = 3.4 T - 4.1 T. (b) $B$ dependence of the resonant frequency $f_0$ extracted from the PlAT (the circles) and $\Delta|S_{21}|$ (the dashed line).}
\end{figure*}

The PlAT can be understood with the Tien-Gordon model \cite{Tien63}, in which oscillating potential is applied across a tunneling barrier. In our case, the circulating plasmon waves with voltage amplitude $\widetilde{V}_{\mathrm{res}}$ in the resonator cause the oscillating potential with amplitude $e\widetilde{V}_{\mathrm{DQD}}$ between the two dots. This oscillating potential splits the eigenstates of the dots with detuning $\varepsilon$ into a superposition of plasmon sidebands with energy $\varepsilon + nhf$ and amplitude $J_n(\alpha)$. The electron in the TLS is dressed with $n$ plasmons in this picture. For the original current profile $I_0(\varepsilon)$ at $\widetilde{V}_{\mathrm{DQD}}$ = 0, the current profile under the oscillating potential can be written as

\begin{equation}
I(\varepsilon)=\sum_n J_n^2(\alpha)I_0(\varepsilon+nhf_0), n = 0, \pm1, \pm2, \ldots,
\end{equation}
where the squared $n^{\mathrm{th}}$-order Bessel function of the first kind $J_n^2(\alpha)$ describes the probability of finding the system in the $n$-plasmon dressed state and thus provides the relative height of the $n$-th satellite peak \cite{Stoof96}. Here, $\alpha = e\widetilde{V}_{\mathrm{DQD}}/hf_0$ is the normalized amplitude of the oscillating potential. By using the Lorentzian profile $I_0(\varepsilon)$ fitted to the data in Fig. 3(a), the observed PlAT current profile can be reproduced with Eq. (1), as shown by the red line in Fig. 3(b) with $\alpha$ = 1.02 ($e\widetilde{V}_{\mathrm{DQD}}$ = 12.4 $\mu$eV).

Multiple-plasmon assisted tunneling is observed at higher microwave powers with the $n$-plasmon peak exhibiting nonlinear power dependence, as shown in Fig. 3(c); the two-plasmon absorption peak at $\varepsilon$ = 24 $\mu$eV and the two-plasmon emission peak at $\varepsilon = -24$ $\mu$eV are seen at $P$ $> -$63 dBm. The first-plasmon peak at $\varepsilon$ = $\pm$12 $\mu$eV shows a maximum at $P = -$57 dBm and the zero-plasmon peak at $\varepsilon$ = 0 vanishes at higher $P = -54$ dBm. These features are reproduced by Eq. (1) with the Bessel function, as shown in Fig. 3(d). All features can be understood with amplitude $J_n(\alpha)$ for $n$-plasmon dressed state. The observation of such a nonlinear optics regime for emergent edge plasmons is remarkable.

To justify that the signal is induced by the edge plasmons, its resonant characteristics are confirmed by measuring the current profiles at various frequencies, as shown in Fig. 4(a). The dashed lines describe the energy quantum $\varepsilon$ = $\pm hf$ for frequency $f$. The PlAT is visible only at around $f_0$ = 2.95 GHz, and no signal is obtained at off-resonant conditions ($f \neq f_0$). The result is well reproduced by our simulation shown in Fig. 4(b), where $\widetilde{V}_{\mathrm{DQD}}$ in Eq. (1) is assumed to be proportional to the Lorentzian profile of $\Delta|S_{21}|$ in Fig. 2(e). The resonant feature safely excludes possible artifacts, for example, associated with the long-range electrostatic crosstalk causing conventional photon-assisted tunneling \cite{Oosterkamp98,Wiel02}.

We repeated similar experiments at several magnetic fields in the range of 3.4 $-$ 4.1 T (corresponding $\nu$ between 2.2 and 1.8), as shown in Fig. 5(a), where the emission peak at $\varepsilon$ $<$ 0 is shown. The resonant frequency $f_0$ changes with $B$, as summarized in Fig. 5(b), in agreement with the $1/B$ dependence (the dashed line) extracted from the resonator characteristics of Fig. 2(c). The satellite peak in Fig. 5(a) is seen in the limited range of $B$ = 3.5 $-$ 4 T in agreement with the data in Fig. 2(c) (see SM \cite{SM}). The result unambiguously ensures that the QH plasmon mode is coupled to the DQD.

Having confirmed the coupling between the DQD and the edge plasmon mode that emerged from electrons, we shall discuss the feasibility of strong coupling regimes of cQED with edge plasmons. The vacuum Rabi splitting $g$ normalized by the plasmon energy $hf_0$ determines the ultrastrong ($\frac{g}{hf_0}$ $\gtrsim$ 0.1) and deep-strong ($\frac{g}{hf_0}$ $\gtrsim$ 1) regimes \cite{Haroche06,Blais21}. For electric dipole coupling, it can be written as $\frac{g}{hf_0} = \frac{1}{2} \zeta_{\mathrm{res}} \eta_{\mathrm{r-q}}$sin$\theta_{\mathrm{qubit}}$ with three dimensionless factors \cite{Childress04}. The resonator factor $\zeta_{\mathrm{res}} = \frac{e\widetilde{V}_{\mathrm{res}}}{hf_0}$ describes the normalized amplitude of the oscillating potential $e\widetilde{V}_{\mathrm{res}}$ of the edge channel for single energy quantum $hf_0$ and is given by $\zeta_{\mathrm{res}} = \sqrt{Z\frac{e^2}{h}}$. Therefore, high-$Z$ resonators are highly desirable for larger $\widetilde{V}_{\mathrm{res}}$ and $\zeta_{\mathrm{res}}$ \cite{Blais04}. The edge plasmon resonator provides high $\zeta_{\mathrm{res}}$ = $\frac{1}{\sqrt{2}}$ at $\nu$ = 2, and $\zeta_{\mathrm{res}} >$ 1 in the fractional QH regimes. The qubit factor sin$\theta_{\mathrm{qubit}}$ = $\frac{2t_{12}}{\sqrt{\varepsilon^2+4t_{12}^2}}$ describes the magnitude of the transition dipole moment, while cos$\theta_{\mathrm{qubit}}$ describes that of the permanent dipole moment \cite{Wiel02}. Whereas the present experiment was done at small sin$\theta_{\mathrm{qubit}} \ll 1$ for demonstrating the PlAT associated with the permanent dipole moment, one can reach sin$\theta_{\mathrm{qubit}}$ $\simeq$ 1 by tuning $t_{12}$ = $\frac{1}{2}hf_0$ and $\varepsilon$ = 0 with gate voltages. Therefore, the coupling factor $\eta_{\mathrm{r-q}}$ = $\frac{\widetilde{V}_{\mathrm{DQD}}}{\widetilde{V}_{\mathrm{res}}}$ needs to be large for the strong coupling regimes.

$\eta_{\mathrm{r-q}}$ can be estimated from our data in the following way. First, we can relate $\widetilde{V}_{\mathrm{res}}$ to the input microwave power $P$ by using the capacitance model shown in Fig. 2(a). The channel can be divided into the injection region of length $L_{\mathrm{inj}}$ with finite capacitance $C_{\mathrm{inj}}$ and the other region of length $L-L_{\mathrm{inj}}$ with $C_{\mathrm{inj}}$ = 0. The edge channel with conductance $\sigma_{xy} = 2e^2/h$ is also capacitively coupled to G$_{\mathrm{res}}$ with $C_{\mathrm{res}}$ and the rest (including the ground) with $C_{\mathrm{ch}}$ and dissipative conductance $G_{\mathrm{s}}$, as shown in Fig. 2(b). The model successfully explains $\widetilde{V}_{\mathrm{res}}$ measured with a quantum dot in our previous study \cite{Ota18}. The width and the height of the resonant peak in $\Delta |S_{21}|$ as well as $\widetilde{V}_{\mathrm{res}}$ can be described by $G_{\mathrm{s}}$ and the product $C_{\mathrm{inj}}L_{\mathrm{inj}}$ in the limit of $L_{\mathrm{inj}} \ll L$. For the data at $P = -$68 dBm in Fig. 2(e), the profile can be fitted with $G_{\mathrm{s}}$ = 0.11 S/m and $C_{\mathrm{inj}}L_{\mathrm{inj}}$ = 1.22 fF, from which we deduce $\widetilde{V}_{\mathrm{res}}$ $\simeq$ 410 $\mu$V and the average number of plasmons $\left \langle n_{\mathrm{pl}}\right \rangle$ $\simeq$ 2200 at the resonant frequency. As we derived $\widetilde{V}_{\mathrm{DQD}}$ = 12.4 $\mu$V at this condition, we estimate $\eta_{\mathrm{r-q}}$ $\simeq$ 0.03. This is reasonably high for the unoptimized device structure. As $\eta_{\mathrm{r-q}}$ strongly depends on the device geometry, we expect to reach $\eta_{\mathrm{r-q}}$ $\gtrsim$ 0.1 by placing the DQD closer to the resonator. Therefore, the ultrastrong regime at $\frac{g}{hf_0}$ $\gtrsim$ 0.1 may be feasible with the scheme potentially.

Another important condition to reach the strong coupling regimes is that the energy loss rates $\kappa$ of the resonator and $\gamma$ of the qubit must be smaller than $g$. If the resonant line widths in the present measurements were determined by these dissipations, we estimate the upper bounds of $\frac{\kappa}{hf_0} \simeq \frac{1}{Q} \simeq 0.06$ and $\frac{\gamma}{hf_0} \simeq \frac{w_{\mathrm{DQD}}}{hf_0} \simeq 0.5$. The strong coupling regimes are expected by improving electrostatic environment of the edge channel for $\kappa$ \cite{Bosco219} and qubit environment for $\gamma$ \cite{Hayashi03,Shinkai09}. These crude estimations encourage us to study cQED with edge plasmons.

In summary, we have demonstrated PlAT in a hybrid quantum system of an edge-plasmon resonator and a DQD. Tunneling with absorption and emission of plasmon quanta manifests quantum properties of emergent plasmons in the quantum Hall regime. The plasmon-electron coupling with the high-impedance plasmon mode is attractive for studying strong coupling regimes of cQED. Overall, our study provides a foundation for combining the plasmons as topological quantum quasiparticles with quantum information devices. This integration makes the system particularly appealing for those new functionalities based on the high-impedance chiral mode.

\vspace{0.3cm}
\begin{acknowledgements} The authors thank T. Hata and W. G. van der Wiel for their beneficial discussions. This study was supported by the Grants-in-Aid for Scientific Research Grants No. JP19H05603, No. JP24K06915, and JST PRESTO Grant No. JPMJPR225C and the TokyoTech-ARIM (Advanced Research Infrastructure for Materials and Nanotechnology) project.
\end{acknowledgements}


\end{document}


\title{Supplementary Information for
`Resonant Plasmon-Assisted Tunnelling in a Double Quantum Dot Coupled to a Quantum-Hall Plasmon Resonator'

}
\author{Chaojing Lin}
\affiliation{Department of Physics, Tokyo Institute of Technology, 2-12-1 Ookayama,
Meguro, Tokyo, 152-8551, Japan.}
\affiliation{JST, PRESTO, 4-1-8 Honcho, Kawaguchi, Saitama 332-0012, Japan.}
\author{Ko Futamata}
\affiliation{Department of Physics, Tokyo Institute of Technology, 2-12-1 Ookayama,
Meguro, Tokyo, 152-8551, Japan.}
\author{Takafumi Akiho}
\affiliation{NTT Basic Research Laboratories, NTT Corporation, 3-1 Morinosato-Wakamiya,
Atsugi, 243-0198, Japan.}
\author{Koji Muraki}
\affiliation{NTT Basic Research Laboratories, NTT Corporation, 3-1 Morinosato-Wakamiya,
Atsugi, 243-0198, Japan.}
\author{Toshimasa Fujisawa}
\affiliation{Department of Physics, Tokyo Institute of Technology, 2-12-1 Ookayama,
Meguro, Tokyo, 152-8551, Japan.}

\date{\today }
\maketitle

\section{1.	Device fabrication and parameters}
Our devices shown in Fig. 1 in the main text were fabricated from a standard GaAs/AlGaAs heterostructure with a two-dimensional electron gas located 100 nm below the surface and having an electron density of $1.8\times10^{11}$ cm$^{-2}$. As a first step of fabrication, the mesa of the wafer was defined by wet chemical etching. After that, ohmic contacts were formed by standard photolithography techniques, followed by the deposition of Au-Ge-Ni metal films and the alloying process. Fine gate structures in the plasmon resonator and the double quantum dot (DQD) were fabricated by using electron-beam lithography techniques and by deposition of Ti/Au (10/20 nm) metal films; In specific, the plasmon resonator was formed by a ring-shaped gate with a radius of 20 $\mu$m and a width of 1 $\mathrm{\mu}$m, and the DQD was defined by several narrow gates with each dot having an approximate size of $\sim100\times100$ nm$^{2}$. Finally, the coplanar microwave transmission lines with a central metal of width 32 $\mu$m and a gap of 18 $\mu$m to the outer metal plane, together with other large gate electrodes for bonding pads were fabricated by using the photolithography and by deposition of Ti/Au (10/250 nm) metal films.

\section{2.	Characteristics of the transmission lines}
\begin{figure}
\includegraphics[width=7. cm]{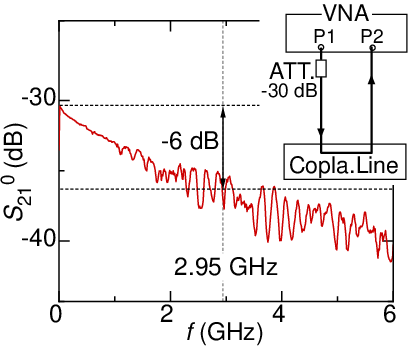}
\caption{Transmission signal $S_{21}^0$ as a function of frequency $f$. The inset shows the measurement setup. An attenuator of $-$30 dB is used in this measurement.}
\label{fig: S1}
\end{figure}

The microwave transmission was characterized by a vector network analyzer (VNA) when the device was cooled at low temperature. The measurement setup is shown in the inset of Fig. S1. The microwave signal from port 1 (P1) of the VNA is introduced to the coplanar transmission line through an attenuator and coaxial cables in a dilution refrigerator. The signal passing through the coplanar line is detected at port2 (P2) of the VNA. Figure S1 shows the transmission spectrum $S_{21}^0$ as a function of frequency $f$, measured without activating the plasmon resonator. The low-frequency data $S_{21}^0$ $\sim -$30.5 dB at $f \sim 0$ is consistent with the 30-dB attenuator. $S_{21}^0$ decreases with increasing the frequency, indicating the loss of coaxial cables and the transmission lines. The total cable loss is 6 dB at the resonant frequency $f_0$ = 2.95 GHz. By assuming identical losses for the forward and return coaxial cables (about 4.5 m for the forward cable and 4 m for the return cables), the microwave power $P$ at the injection gate $G_{\mathrm{inj}}$ should be 33 dB smaller than the incident microwave power $P_{\mathrm{in}}$ ($P$ = $P_{\mathrm{in}}$ $-$ 33 dB).

The $S_{21}^0$ spectrum shows oscillations with a period $\Delta f \sim$ 0.24 GHz, which may be associated with spurious resonances in the cables. These oscillations with modulation of $\pm 1$ dB are effectively removed by subtracting the data $S_{21}^0$ without activating the resonator from the data $S_{21}$ taken with the resonator ($\Delta S_{21} = S_{21} - S_{21}^0$) for the plots in Fig. 2.

For the measurements on plasmon assisted tunneling, we replaced the VNA with a microwave generator for P1 and a 50-$\Omega$ terminator for P2. The same attenuator-cable loss (33 dB in total) is assumed. The dc current $I_{\mathrm{d}}$ through the DQD was measured under continuous wave irradiation, and no subtraction scheme is employed. While the oscillations associated with spurious resonances are expected, they did not appear in the current spectra of Fig. 4(a) and Fig. 5(a), partly because the resonance width (0.16 GHz) of the plasmon resonator was narrower than $\Delta f \sim$ 0.24 GHz.

\section{3.	Distributed circuit model of the plasmon resonator}
\begin{figure}
\includegraphics[width=12 cm]{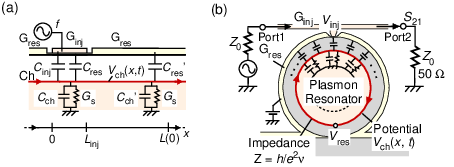}
\caption{(a) Distributed circuit model for the plasmon resonator. (b) Measurement setup for the plasmon resonator.}
\end{figure}

To obtain a quantitative understanding of the resonance spectra, we developed a distributed circuit model for the plasmon resonator. Generally, charge density $\rho_{\mathrm{ch}}$ and the voltage $V_{\mathrm{ch}}$ of the edge channel are related by $\rho_{\mathrm{ch}} = C_{\mathrm{\Sigma}}V_{\mathrm{ch}}$ with a proportional factor $C_{\mathrm{\Sigma}}$ (capacitance per unit length). For convenience, $C_{\mathrm{\Sigma}}$ can be decomposed into capacitances $C_{\mathrm{res}}$ for the gate G$_{\mathrm{res}}$ defining the resonator, $C_{\mathrm{inj}}$ for the injection gate G$_{\mathrm{inj}}$, and $C_{\mathrm{ch}}$ for the rest including the QH bulk insulator. For the gate geometry shown in the SEM image (the upper right inset) of Fig. 2(a), the capacitances ($C_{\mathrm{res}} + C_{\mathrm{ing}} +C_{\mathrm{ch}} = C_{\mathrm{\Sigma}}$) in the injector segment of length $L_{\mathrm{inj}}$ = 1 $\mu$m can be different from those ($C_{\mathrm{res}}' + C_{\mathrm{ch}}' = C_{\mathrm{\Sigma}}'$ and $C_{\mathrm{inj}}' = 0$) in the main segment of length $L - L_{\mathrm{inj}}$. As our electrostatic simulation suggests $C_{\mathrm{\Sigma}} \cong C_{\mathrm{\Sigma}}'$ and $C_{\mathrm{res}} + C_{\mathrm{inj}} \cong C_{\mathrm{res}}'$, we assume $C_{\mathrm{inj}} \leq C_{\mathrm{res}} = C_{\mathrm{res}}'$ for simplicity.

Application of time-dependent microwave voltage $V_{\mathrm{inj}}(t)$ to the injection gate induces a displacement current density, $C_{\mathrm{inj}}\mathrm{d}(V_{\mathrm{inj}}-V_{\mathrm{ch}})/\mathrm{d}t$, in the channel. This excites a plasmon wave with $V_{\mathrm{ch}}(x, t)$. The electrical current of the edge channel is $I_{\mathrm{ch}} = \sigma _{\mathrm{xy}}V_{\mathrm{ch}}$, where $\sigma _{\mathrm{xy}}$ (= $Z^{-1}$) is the Hall conductance. The decay of plasmons in the channel is accounted for by introducing a scattering conductance $G_{\mathrm{s}}$ between the channel and the ground. The current conservation law provides the wave equation in a form \cite{Ota18},

\begin{equation}
C_{\mathrm{\Sigma}}\frac{\mathrm{d}V_{\mathrm{ch}}}{\mathrm{d}t} = -\sigma _{\mathrm{xy}}\frac{\mathrm{d}V_{\mathrm{ch}}}{\mathrm{d}x} + C_{\mathrm{inj}}\frac{\mathrm{d}V_{\mathrm{inj}}}{\mathrm{d}t} - G_{\mathrm{s}}V_{\mathrm{ch}}, [0 \leq x \leq L_{\mathrm{inj}}], \label{S1}
\end{equation}
\begin{equation}
C_{\mathrm{\Sigma}}\frac{\mathrm{d}V_{\mathrm{ch}}}{\mathrm{d}t} = -\sigma _{\mathrm{xy}}\frac{\mathrm{d}V_{\mathrm{ch}}}{\mathrm{d}x} - G_{\mathrm{s}}V_{\mathrm{ch}}, [L_{\mathrm{inj}} \leq x \leq L], \label{S2}
\end{equation}

We solve these equations by using periodic boundary conditions at $x$ = 0 and $L_{\mathrm{inj}}$, For an excitation voltage of the form, $V_{\mathrm{inj}}(t) = \widetilde{V}_{\mathrm{inj}}e^{i\omega t}$ with complex amplitude $\widetilde{V}_{\mathrm{inj}}$, on G$_{\mathrm{inj}}$, the complex amplitude of the potential of the channel is given by

\begin{equation}
\widetilde{V}_{\mathrm{ch}} = \frac{i\omega C_{\mathrm{inj}}}{G_{\mathrm{s}}+i\omega C_{\Sigma}}[1-\frac{1-e^{-ik(L-L_{\mathrm{inj}})}}{1-e^{-ikL}}e^{-ikx}]\widetilde{V}_{\mathrm{inj}}, [0 \leq x \leq L_{\mathrm{inj}}], \label{S3}
\end{equation}
\begin{equation}
\widetilde{V}_{\mathrm{ch}} = \frac{i\omega C_{\mathrm{inj}}}{G_{\mathrm{s}}+i\omega C_{\Sigma}}[\frac{1-e^{-ikL_{\mathrm{inj}}}}{1-e^{-ikL}}e^{-ik(x-L_{\mathrm{inj}})}]\widetilde{V}_{\mathrm{inj}}, [L_{\mathrm{inj}} \leq x \leq L], \label{S4}
\end{equation}
where $k = \omega/v_{\mathrm{p}} + iG_{\mathrm{s}}/\sigma_{\mathrm{xy}}$ is the complex wave number with the plasmon velocity $v_{\mathrm{p}} = \sigma_{\mathrm{xy}}/C_{\Sigma}$ and the plasmon decay rate $G_{\mathrm{s}}/\sigma_{\mathrm{xy}}$. The factor $\frac{1}{1-e^{-ikL}}$ in Eqs. (S3) and (S4) determines the resonant frequency $\omega_0 = 2\pi \sigma_{\mathrm{xy}}/LC_{\Sigma}$.

For the measurement circuit shown in Fig. S2(b), the plasmon resonator can be characterized by an effective impedance $Z_{\mathrm{res}} = \widetilde{V}_{\mathrm{inj}}/\widetilde{I}_{\mathrm{inj}}$, where $\widetilde{I}_{\mathrm{inj}} = \int_{0}^{L_{\mathrm{inj}}} i\omega C_{\mathrm{inj}}(\widetilde{V}_{\mathrm{inj}}-\widetilde{V}_{\mathrm{ch}})\mathrm{d}x$ is the complex amplitude of the displacement current flowing through the injection gate G$_\mathrm{inj}$. The resultant $Z_{\mathrm{res}}$ can be expressed as

\begin{equation}
Z_{\mathrm{res}}^{-1} = \omega C_{\mathrm{inj}}L_{\mathrm{inj}} \frac{i\omega C_{\mathrm{inj}}}{G_{\mathrm{s}} + i\omega C_{\Sigma}}[\frac{1-e^{-ikL_{\mathrm{inj}}}}{kL_{\mathrm{inj}}}\frac{1-e^{-ik(L-L_{\mathrm{inj}})}}{1-e^{-ikL}} + i(\frac{C_{\Sigma}}{C_{\mathrm{inj}}}-1+\frac{G_{\mathrm{s}}}{i\omega C_{\mathrm{inj}}})], \label{S5}
\end{equation}
With this impedance, the transmission signal $S_{21}$ from port 1 to port 2 is then given as

\begin{equation}
S_{21}(\omega) = \frac{2Z_{\mathrm{res}}}{2Z_{\mathrm{res}}+Z_0}, \label{S6}
\end{equation}
Based on this equation, $S_{21}^0 = 1$ (0 dB) can be obtained by setting $Z_{\mathrm{res}} = \infty$ for an open load with no resonator being formed.

As Eq. (S6) with Eq. (S5) is complicated to analyze the resonance characteristics, we use the approximations of $L_{\mathrm{inj}} \ll L$ for small injection and $G_{\mathrm{s}}L \ll \sigma_{\mathrm{xy}}$ for weak decay to simplify the expression. The spectra around the resonance ($\omega \sim \omega_0$) then become,

\begin{equation}
S_{21}(\omega) = 1-\frac{Z_0\sigma_{\mathrm{xy}}^2}{2(LC_{\mathrm{\Sigma}})^2}\frac{(2\pi C_{\mathrm{inj}}L_{\mathrm{inj}})^2\omega}{\omega_0(G_{\mathrm{s}}L)+2\pi \sigma_{\mathrm{xy}}i(\omega-\omega_0)}, \label{S7}
\end{equation}
We use this equation to fit the resonance peak of $\Delta |S_{21}|$ in Fig. 2(e) in the main text. From the fitting, we obtain two values: $C_{\mathrm{inj}}L_{\mathrm{inj}}$ = 1.22 fF and $G_{\mathrm{s}}$ = 0.11 S/m (or $G_{\mathrm{s}}L$ = 13.86 $\mu$S). The first value describes the plasmon injection efficiency through the gate G$_{\mathrm{inj}}$, and the second value associated with the peak width describes the plasmon decay rate.

\begin{figure}[t]
\includegraphics[width=9 cm]{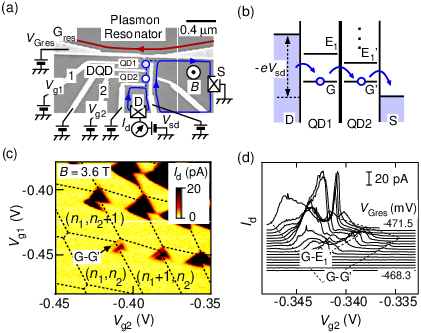}
\caption{(a) An SEM image of the device with the measurement setup for the DQD. (b) Transport through the DQD with multiple energy levels. (c) Tunnel current $I_{\mathrm{d}}$ as a function of gate voltage $V_{\mathrm{g1}}$ and $V_{\mathrm{g2}}$. (d) Waterfall plot of $I_{\mathrm{d}}$. The bias triangles are measured by varying $V_{\mathrm{g2}}$ and $V_{\mathrm{Gres}}$, while $V_{\mathrm{g1}}$ is fixed at $-$220 mV. Sharp resonance peaks can be obtained with this measurement scheme.}
\end{figure}

As shown in Fig. S2(b), the DQD is effectively coupled to the resonator at $x_{\mathrm{c}} = (L + L_{\mathrm{inj}})/2$. Equation (S4) suggests $\widetilde{V}_{\mathrm{res}}/\widetilde{V}_{\mathrm{inj}} \cong \omega_0(C_{\mathrm{inj}}L_{\mathrm{inj}}/G_{\mathrm{s}}L) \cong 1.6$ at $f_0$ = 2.95 GHz under the above approximations. This value is used to estimate the coupling factor $\eta_{\mathrm{r-q}}$, the amplitude $|\widetilde{V}_{\mathrm{res}}|$, and the average number of plasmons $\langle n_{\mathrm{pl}} \rangle = \frac{|\widetilde{V}_{\mathrm{res}}|^2/Z}{hf_0^2}$ in the main text.

\section{4.	Sample and transport characteristics of the DQD}

Figure S3(a) shows an SEM image of a DQD device in the vicinity of the plasmon resonator. The ring-shaped gate G$_{\mathrm{res}}$ is used to form a plasmon resonator, and the five finger gates are used to define the DQD marked by two blue circles. The tunneling current $I_{\mathrm{d}}$ through the DQD is measured with a bias voltage $V_{\mathrm{sd}}$ (= 400 $\mu$V) applied between the drain (D) and the source (S). The energy levels in the dot 1 (dot 2) are tuned by applying the gate voltage $V_{\mathrm{g1}}$ ($V_{\mathrm{d2}}$) to gate 1 (2). The quantum Hall state at filling factor $\nu$ is set by applying a perpendicular magnetic field $B$ to the device.

Figure S3(c) shows the stability diagram of the DQD measured at $B$ = 3.6 T ($\nu \sim$ 2) in the absence of microwave irradiation ($P$ = 0). The measured $I_{\mathrm{d}}$ is plotted in the color scale as a function of the gate voltages $V_{\mathrm{g1}}$ and $V_{\mathrm{g2}}$ at $V_{\mathrm{Gres}} = -$0.334 V. Triangular conductive regions are located at the corners of honeycomb-shaped stability diagram represented by the dashed lines \cite{van02}. The size of the triangles measures the transport window of $eV_{\mathrm{sd}}$ = 400 $\mu$eV, from which the lever arm factor was determined. The spacings between triangles measures the Coulomb charging energy of $\sim$0.8 meV for both dots and the interdot electrostatic coupling energy of $\sim$86 $\mu$eV. Some triangles show fine peaks associated with the resonant tunneling between the discrete energy levels of the two dots, from which typical single-particle energy spacing of $\sim$140 $\mu$eV is estimated for the two dots.

To resolve plasmon-induced tunneling, it is crucial to obtain a sharp tunneling peak with a peak width much smaller than the plasmon energy $hf$. Figure S3(d) shows a waterfall plot of the current spectra for a particular triangle region measured by varying the gate voltages $V_{\mathrm{g2}}$ and $V_{\mathrm{Gres}}$ while $V_{\mathrm{g1}}$ was fixed at $-$220 mV. Here, we used $V_{\mathrm{Gres}}$ to control the energy level of dot 1 to control the tunneling probability between dot 1 and the source (S). While the peaks labeled G-G' show small current, we found a reasonably sharp peak labeled G-E1', corresponding to resonant tunneling through the ground state G in dot 1 and the excited state E1' in dot 2. We used this peak to investigate the plasmon assisted tunneling.

\section{5.	Tunneling current spectra with and without plasmon irradiation}
\begin{figure}[b]
\includegraphics[width=14 cm]{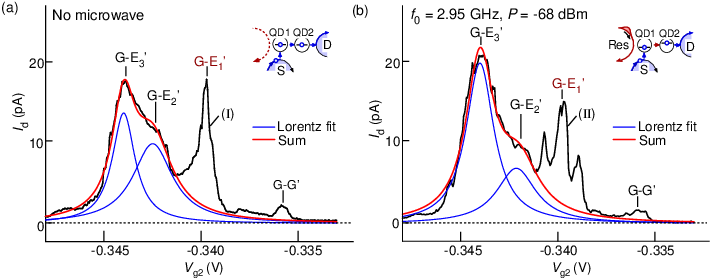}
\caption{(a-b) Tunneling current $I_{\mathrm{d}}$ as a function of $V_{\mathrm{g2}}$ measured (a) without microwave power and (b) with microwave applied at $f$ = 2.95 GHz and $P = -$68 dBm. Several resonant tunneling peaks are observed. The blue lines are Lorentz fits to peaks G-E2' and G-E3'. Their sum is shown as the red line and used to remove the background current for the peak G-E1'.}
\end{figure}

Figure S4(a) presents the current trace $I_{\mathrm{d}}$ as a function of $V_{\mathrm{g2}}$, obtained with $V_{\mathrm{g1}} = -$220 mV and $V_{\mathrm{Gres}} = -$469.6 mV and no microwave power. Several resonant tunneling peaks are observed, which are labeled as G-G', G-E1', G-E2', ... in the figure \cite{Vaart95}. A particularly sharp peak is found at G-E1'. This peak is replotted as the trace (I) in Fig. 2(a) in the main text, where the gate voltage in the horizontal axis has been converted to the detuning energy of the double dot by using a lever-arm factor (13.3 $\mu$eV/mV). To remove the background current for peak G-E1', we first fit the peaks G-E2', G-E3' using the Lorentzian function (blue lines) and then subtract their sum (red line) from the experimental trace. The result is plotted as the trace (I') in Fig. 2(a) in the main text.

Under plasmon irradiation, in addition to the main peak (G-E1'), distinct satellite peaks are observed in the current trace $I_{\mathrm{d}}$ in Fig. S4(b), measured at $f$ = 2.95 GHz and $P = -$68 dBm. The left and right peaks are associated with the absorption and emission of a single plasmon, respectively \cite{Stoof96}. These peaks are plotted as the trace (II) in Fig. 2(b) in the main text with the horizontal axis converted to the detuning energy in the quantum dot. The satellite peaks are not resolved for other tunneling peaks due to their broad peak width ($> hf$). Using similar fitting and subtraction procedures, the background current of peak G-E1' can be removed, and the result is plotted as the trace (II') in Fig. 2(b) in the main text.

\section{6.	Comparison between the satellite peak value and the plasmon power}
\begin{figure}
\includegraphics[width=7. cm]{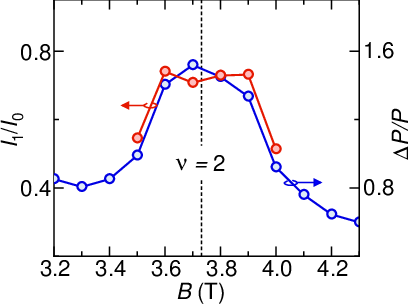}
\caption{Satellite peaks $I_1$, normalized with the main peak $I_0$, as $I_1/I_0$ is plotted a function of $B$ on the left axis. The data is extracted from Fig. 5(a) in the main text. The plasmon power $\Delta P$ normalized by injection power $P$, as $\Delta P/P$ is shown on the right axis. This data is calculated from the spectra $\Delta |S_{21}|$ in Fig. 2(c) in the main text.}
\end{figure}

The plasmon-assisted tunneling was studied under different magnetic fields. The measured current tunneling spectra were presented in Fig. 5(a) in the main text. Here, we summarize the magnetic field $B$ dependence of the peak value $I_1$, which is plotted on the left axis of Fig. S5 with its value normalized by the main peak $I_0$. Large $I_1/I_0$ values are observed in the $B$ range near $\nu$ = 2, while the value decreases when $\nu$ deviates from 2. This change can be compared to the plasmon transport properties in the resonator. We extracted the plasmon power $\Delta P/P = 1-10^{\Delta |S_{21}|/10}$ from the spectra $\Delta |S_{21}|$ in Fig. 2(c) in the main text. The results were plotted on the right axis of Fig. S5, which is shown as a linear scale for a direct comparison with $I_1/I_0$. Here, $\Delta P/P$ characterizes the relative power reduction at the VNA detection port P2 when the resonator is activated. According to Eq. (1) in the main text, the ratio $I_1/I_0 \sim J_{1}^2(\alpha)$ can be related to the Bessel function $J_n^2(\alpha)$ with $n$ = 1 for 1-plasmon assisted tunneling. In the small limit of $\alpha \ll 1$, $J_1^2(\alpha) \sim \alpha^2 \sim V_{\mathrm{DQD}}^2$ can be further approximated. Thus, $I_1/I_0$ is expected to be proportional to the plasmon power in the resonator, which is indeed seen in Fig. S5. The similarity between the two data sets further confirms the coupling of plasmons and the DQD. This result also clearly demonstrates the significant impact of the quantum Hall regime on plasmon transport. As generally plasmon dissipation increases with local disorders scattering around the channel or in the bulk \cite{Ashoori92}, better insulating bulk states with well-defined quantum Hall regimes are therefore desirable for obtaining a good plasmon resonance.
